\documentclass[sigconf]{acmart}
\usepackage{enumitem}
\AtBeginDocument{%
  }

\copyrightyear{2025}
\acmYear{2025}
\setcopyright{rightsretained}
\acmConference[RecSys '25]{Proceedings of the Nineteenth ACM Conference on Recommender Systems}{September 22--26, 2025}{Prague, Czech Republic}
\acmBooktitle{Proceedings of the Nineteenth ACM Conference on Recommender Systems (RecSys '25), September 22--26, 2025, Prague, Czech Republic}\acmDOI{10.1145/3705328.3748118}
\acmISBN{979-8-4007-1364-4/2025/09}

\begin{document}

\title{RankGraph: Unified Heterogeneous Graph Learning for Cross-Domain Recommendation}

\author{Renzhi Wu}
\affiliation{%
  \institution{Meta MRS}
  \city{Menlo Park}
  \country{US}}
\email{renzhiwu@meta.com}
\author{Junjie Yang}
\affiliation{%
 \institution{Meta MRS}
  \city{Menlo Park}
  \country{US}}
\email{junjieyang@meta.com}
\author{Li Chen}
\affiliation{%
  \institution{Meta MRS}
  \city{Menlo Park}
  \country{US}}
\email{lichenntu@meta.com}

\author{Hong Li}
\affiliation{%
  \institution{Meta MRS}
  \city{Menlo Park}
  \country{US}}
\email{hongli@meta.com}

\author{Li Yu}
\affiliation{%
  \institution{Facebook Monetization}
  \city{Menlo Park}
  \country{US}}
\email{yongnali@meta.com}

\author{Hong Yan}
\affiliation{%
  \institution{Meta MRS}
  \city{Menlo Park}
  \country{US}}
\email{hyan@meta.com}


\renewcommand{\authors}{Renzhi Wu, Junjie Yang, Li Chen, Hong Li, Li Yu, Hong Yan}
\begin{abstract}
 Cross-domain recommendation systems face the challenge of integrating fine-grained user and item relationships across various product domains. To address this, we introduce RankGraph, a scalable graph learning framework designed to serve as a core component in recommendation foundation models (FMs). By constructing and leveraging graphs composed of heterogeneous nodes and edges across multiple products, RankGraph enables the integration of complex relationships between users, posts, ads, and other entities. Our framework employs a GPU-accelerated Graph Neural Network and contrastive learning, allowing for dynamic extraction of subgraphs such as item-item and user-user graphs to support similarity-based retrieval and real-time clustering. Furthermore, RankGraph integrates graph-based pretrained representations as contextual tokens into FM sequence models, enriching them with structured relational knowledge. 
 RankGraph has demonstrated improvements in click (+0.92\%) and conversion rates (+2.82\%) in online A/B tests, showcasing its effectiveness in cross-domain recommendation scenarios.
\end{abstract}




\maketitle

\section{Introduction}
Graph learning has emerged as a powerful approach in recommendation systems, leveraging the rich structure of graphs to model complex relationships between entities. Unlike traditional models that focus on per user’s independent data points, graph learning treats data as interconnected nodes and edges, enabling more natural and effective representations of real-world phenomena. With its ability to capture dependencies, interactions, and contextual information, graph learning is widely adopted in Recommendation Systems~\cite{wang2021graph, wu2022graph}.

In recent years, the concept of recommendation foundation models (FMs) has gained considerable momentum~\cite{zhai2024actions, damianou2024towards, liu2023towards}. These models serve as backbones that can be adapted to a wide range of downstream tasks. However, integrating fine-grained user and item relationships across various product domains remains a significant challenge in cross-domain recommendation FMs~\cite{zhu2021cross, zang2022survey}.

To address this challenge, we introduce RankGraph, a unified graph learning framework designed to harness the power of graph learning within the context of recommendation FMs.  By constructing and leveraging graphs composed of heterogeneous nodes and edges across multiple products, RankGraph enables the integration of complex relationships between users, posts, ads, and other entities. RankGraph demonstrates how graph structures can serve as a core component in enhancing the capabilities of recommendation foundation models, opening new avenues for cross-domain improvements.

\section{RankGraph System Architecture}

\begin{figure*}[h]
  \centering
  \includegraphics[width=\textwidth]{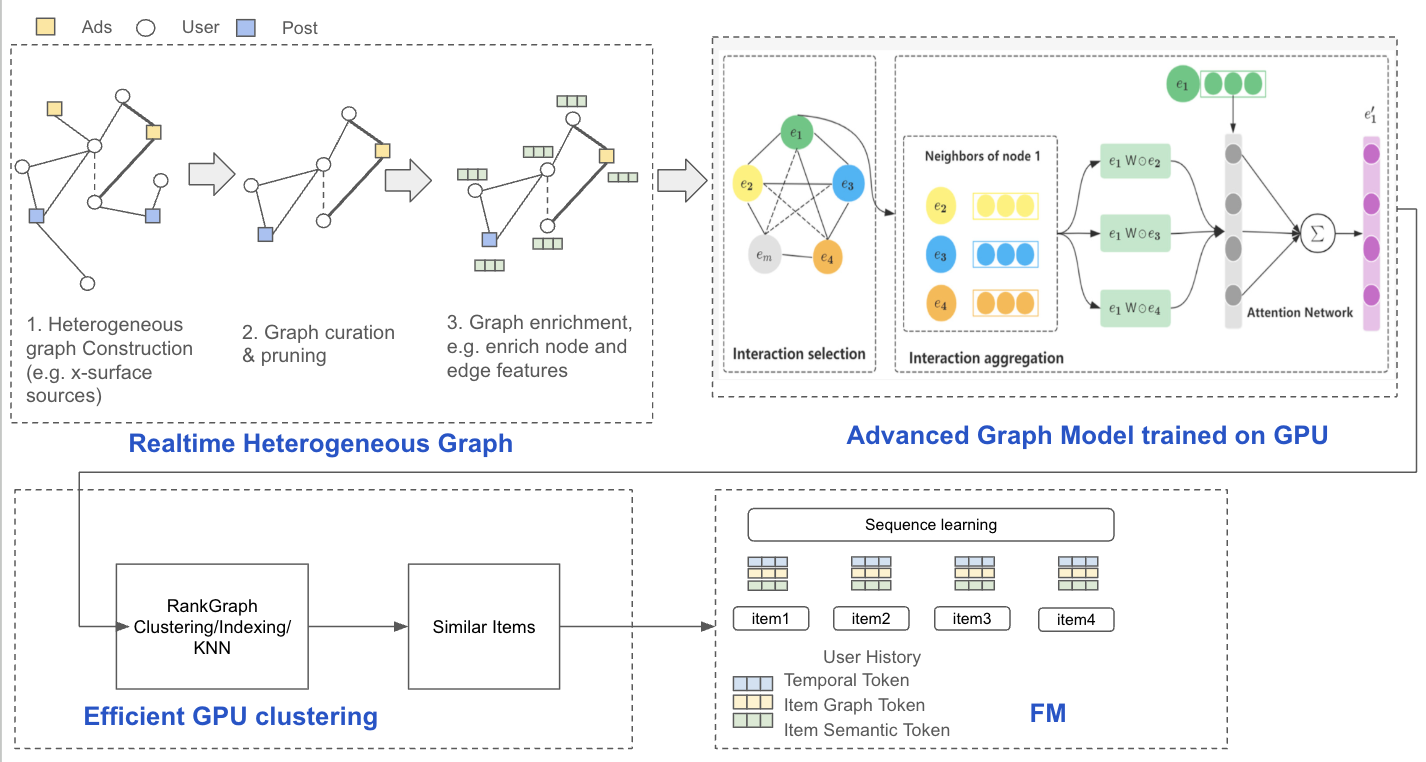}
  \caption{Rankgraph System Architecture.}
  \label{fig:rankgraph}
\end{figure*}

The high level architecture of RankGraph is shown in Figure~\ref{fig:rankgraph}.

\subsection{Heterogeneous Graph}
RankGraph constructs a comprehensive heterogeneous graph to capture the complex, multi-relational nature of recommendation systems. The graph consist of diverse node and edge types derived from cross-product interactions.
Edges encode engagement signals using weighted combinations of different interaction types (e.g., clicks, likes, shares). 

In addition to direct engagement edges captured in the adjacency matrix, \textit{semantic edges} are introduced to model higher-order relationships. These edges represent indirect interactions through multi-hop neighbors, allowing the graph to capture richer contextual and behavioral semantics.

This heterogeneous structure enables RankGraph to effectively model cross-surface foundation model (FM) scenarios, which are difficult to represent using homogeneous graphs. However, the added heterogeneity also brings challenges in semantic representation and learning, addressed through a specialized GPU-based Graph Neural Network described in the next section.

\subsection{Model Architecture}
To learn high quality representations from heterogeneous graphs, RankGraph incorporates several 
components. 

\subsubsection{Graph Feature Encoder}
Node features differ across types and lie in disparate feature spaces. Each node type $t$ contains $n_t$ different feature types (e.g. raw id features and semantic embedding from other models). For each node type, RankGraph projects these features into a unified embedding space using the transformation:
\[
h_t = M_t(\text{concat}_{j=1}^{n_t}f_{t,j}(x_{t,j}))
\]
where $x_{t,j}$ is the feature matrix for node type $t$ and its $j$th feature type, $f_{t,j}$ is an MLP; $\text{concat}_{j=1}^{n_t}$ denotes concatenating the output from each $f_{t,j}$; $M_t$ is a feature mixer that combines all feature types and their interactions ( i.e. multiplication of each two feature types). 


\subsubsection{Information Aggregation (RGCN-style)}
RankGraph adopts a message-passing mechanism inspired by Relational Graph Convolutional Networks (RGCN)~\cite{schlichtkrull2018modeling}. The node update rule for relation $r$ at layer $l+1$ is:
\[
h_i^{l+1} = M_t\left(\text{concat}_{r} (f_r(c_{i,r} \sum_{j \in \mathcal{N}^r_i} W_r h_j^{l}))  \right) 
\]
where $\mathcal{N}^r_i$ denotes the set of neighbors of node $i$ under relation $r$ (there are many relations, e.g. click relation between user node and ad node, and co-engagement relation between two ads), $c_{i,r}$ is a normalization factor, $W_r$ is a relation-specific weight matrix, and $h_j^l$ is the projected features from the previous layer; $M_t$ is a feature mixer that combines the aggregated embedding of each relation. This formulation enables each node to aggregate contextual information from its neighbors while preserving original features via self-loops.

\subsubsection{Contrastive Learning}
To improve representation quality, RankGraph employs contrastive learning, training the model to distinguish between positive node pairs (with existing edges) and negative pairs (without edges). This encourages semantically similar nodes to have similar embeddings, enhancing downstream performance. This step involves negative sampling and designing the contrastive loss. 

\begin{itemize}
     
\item\textbf{Negative sampling.} RankGraph employs negative sampling to construct negative pairs, utilizing three distinct methods: (1) in-batch sampling. This is to sample nodes from different edges within the same training batch as negatives. (2) out-of-batch sampling. This approach samples nodes across different batches to ensure that the distribution of negatives is closer to the global distribution. To achieve efficient out-of-batch sampling, we maintain a candidate pool on GPU for each node type and incrementally update the pool after loading each training batch. (3) semantic negative sampling. The model components (such as feature encoders and aggregators) are designed with multiple heads. For each negative example pair, embeddings generated by different heads are used as additional negative examples, enhancing the robustness of the model.

\item\textbf{Contrastive loss.}
RankGraph uses a combination of the triplet loss~\cite{schroff2015facenet, dong2018triplet} and the infonce loss~\cite{oord2018representation, parulekar2023infonce}. The underlying intuition behind this combination is that the triplet loss separates individual positive and negative pairs at a local level, whereas the infonce loss operates on a global scale, distinguishing between clusters of positive and negative pairs. This synergy enables RankGraph to capture both local and global relationships within the data, leading to improved performance.
\end{itemize}

\subsection{Real-Time Training and Serving}
RankGraph is built for real-time operation, with both training and inference fully GPU-accelerated. Once trained, the model can generate node embeddings on-the-fly for high-throughput recommendation scenarios. Additionally, we implement GPU-optimized node clustering algorithms for efficient identification of similar items or users in large-scale graphs. This allows rapid retrieval of similar posts, ads, or users based on recent interactions. RankGraph has been 
launched as a retrieval generator, 
improving the overall user experience through personalized recommendations.

Furthermore, RankGraph offers the capability to extract diverse subgraphs, including \textit{item-item} and \textit{user-user} graphs, from the primary heterogeneous graph. This feature enables use cases where these subgraphs are sufficient, providing flexibility and adaptability in various applications.

\subsection{Integration with Foundation Models}
The resulting graph embeddings can be integrated as input tokens into sequence-based foundation models. These \textit{graph tokens} are combined with other token types (e.g., timestamp tokens) to enhance the expressiveness of user-to-item recommendation models. This allows RankGraph to inject structured graph knowledge into FM pipelines, improving personalization and ranking performance across surfaces.

\section{Experiments}

In this section, we present the evaluation results to demonstrate the effectiveness of the RankGraph system. 

\subsection{Offline Evaluation.} We investigate the quality of item embeddings generated by RankGraph through offline recall evaluation. We use Filament2, a commonly used graph learning system at Meta as our baseline.

\subsubsection{Eval Recall on Next-day Graph Edges}
We evaluate the performance of the methods using the embedding generated in the previous day to evaluate the recall on the graph edges generated in the next day. Specifically, we randomly sample 1000 edges $E_s$ and for the nodes from the sampled edges we compute the distances of every pair of them. For each node $i$, the other node $j$ would be a ground-truth positive example if edge $(i, j) \in E_s$. For each node, we calculate the recall of its ground-truth positive example in  its top $k$ neighbors, and report the averaged recall over all nodes in Table~\ref{tab:eval_recall}.

\begin{table}[ht!]
\caption{Eval Recall for RankGraph and Filament2 on 1000 sampled edges}
\label{tab:eval_recall}
\begin{tabular}{@{}lllll@{}}
\toprule
Method    & Recall@5 &Recall@10 &Recall@50 &Recall@100  \\ \midrule
Filament2  & 0.051 &0.079 & 0.268 & 0.379   \\
RankGraph & 0.143 &0.239 & 0.485 & 0.614 
\end{tabular}
\end{table}

\subsubsection{Engagement Recall}
The recall metrics in Table~\ref{tab:eval_recall} measures how good the methods fit the graph dataset. However, these recall metrics not necessarily present the power of the embeddings in predicting user's future engagement, and in fact we have observed discrepancy between these offline recall metrics and online a/b test results. 
Therefore, to reduce the discrepancy, we proposed a \textit{engagement recall} metric, calculated with the following procedure:
\begin{itemize}
    \item At hour $t$, we get the latest item embeddings generated by RankGraph and obtain the top 20 nearest neighbors for each item. 
    \item For each user, we use the interacted items in the past week as triggers (weighted by the type of interaction), and their the nearest neighbors as the predicted items that each user may interact in the future. The predicted items are sorted by trigger weight and the embedding similarity score with the corresponding trigger. 
    \item We use the ground-truth interactions between hour $t+1$ and hour $t+4$ to calculate the recall metrics of our generated predictions. 
\end{itemize}
We report the recall metric averaged over one day on data from a product surface with billions of users in Table~\ref{tab:recall}. 

\begin{table}[ht!]
\caption{Engagement Recall for RankGraph and Filament2}
\label{tab:recall}
\begin{tabular}{@{}llll@{}}
\toprule
Method    & Recall@100 &Recall@200 &Recall@500 \\ \midrule
Filament2  & 0.071 &0.125 & 0.221    \\
RankGraph & 0.106 &0.157 & 0.234  
\end{tabular}
\end{table}



\section{Conclusion}
In this work, we introduced RankGraph, a scalable and efficient graph-based framework serve as a core component in recommendation foundation models (FMs). By constructing heterogeneous graphs that incorporate multi-type nodes and semantic edges, RankGraph captures the complex, multi-relational nature of user-item interactions across surfaces. The proposed GPU-accelerated Graph Neural Network architecture includes a type-aware feature encoder, relational message passing, and contrastive learning, enabling the model to learn robust, high-quality node embeddings.  



\section{Speaker Bio}
\textbf{Hong Li} is a research scientist director from Meta MRS working on improving Meta recommendation system.

\begin{acks}
This work would not be possible without work from the following
contributors: Hang Wang, Jeff Wang, Honghao Wei, Yiyi Pan, Yinglong Xia, Jason Liu, Hanqing Zeng, Gilbert Jiang, Ren Chen, Hunter Song, Yao Zhang, Arthi Suresh, Jiang Li, Yinglong Xia, Jason Liu, Hao Lin, Siqi Yan, Yanzun Huang, Hao Wang, Wei Zhao, Liang Zhang, Yuming Liu, Hongming Pu, Harrison Zhao, Ziyi Zhao, Lei Huang, Harry Hai Nguyen, Anand Iyer, Jim Li, Tao Ju, Crystal Jin, Ye Wang, Mingda Li, Cong Zhang, Jun Seok Lee, Zhen Wang, Tian Tong, Prabhjot Singh, Pu Zhang, Keke Zhai, Lillian Zhang, Jiazhou Wang,  Emy Sun, Lei Chen, Xiaoxing Zhu, Yuting Zhang, Zhe (Joe) Wang, Daisy Shi He, Min Ni, Bi Xue, Sophia (Xueyao) Liang, Yang Cao, Chengye Liu, Pan Chen, Jiacheng Wu, Yucheng Zhang, Shang Huang, Jun Xiao, Max Fan, Lu Zhang, Xinyao Hu, Shilin Ding, Haomin Yu, Ke Pan, Jianhui Wu, Yuanyuan Ding, Haoran Wen, Serena Li, Lizhu Zhang, Jack Zhai, Ke Gong, Rohan Katpelly, Ram Ramanathan, Nipun Mathur, Helen Ma, Deepak Vijaywargi, Divij Pasrija
\end{acks}

\bibliographystyle{ACM-Reference-Format}
\bibliography{sample-base}


\end{document}